OPEN

# Co-evaporation as an optimal technique towards compact methylammonium bismuth iodide layers

Cristina Momblona, Hiroyuki Kanda, Albertus Adrian Sutanto, Mounir Mensi, Cristina Roldán-Carmona✉ & Mohammad Khaja Nazeeruddin✉

**The most studied perovskite-based solar cells reported up to date contain the toxic lead in its composition. Photovoltaic research and development towards non-toxic, lead-free perovskite solar cells are critical to finding alternatives to reduce human health concerns associated with them. Bismuth-based perovskite variants, especially in the form of methylammonium bismuth iodide (MBI), is a good candidate for the non-toxic light absorber. However, the reported perovskite variant MBI thin films prepared by the solution process so far suffers from poor morphology and surface coverage. In this work, we investigate for the first time the optoelectronic, crystallographic and morphological properties of MBI thin films prepared via thermal co-evaporation of MAI and $BiI_3$. We find by modifying the precursor ratio that the layer with pure MBI composition lead to uniform, compact and homogeneous layers, broadening the options of deposition techniques for lead-free based perovskite solar cells.**

Organic–inorganic metal halide perovskites have emerged as promising candidates for the next generation of photovoltaics (PV) given their unprecedented rise in power conversion efficiency (PCE) achieving 25.2%[1] Beyond their superior optoelectronic properties, low cost and high versatility[2–5], their instability and lead toxicity are at present major concerns, which greatly limit their incorporation into marketable products. Significant efforts are therefore oriented into the search of stable and non-toxic perovskite materials, yet real alternatives based on lead-free components are still missing.

Among the various lead-free perovskites variants proposed in the literature, Sn-based materials have demonstrated promising efficiencies leading to a PCE value of 9.6%[6]. However, Sn-based perovskites suffer from inevitable oxidation of $Sn^{2+}$ to $Sn^{4+}$, providing a source of inhomogeneity and pinholes that are further enhanced by their fast crystallization[7]. As a result, the most straightforward candidates possessing similar electronic properties to $Pb^{2+}$, which are suitable for light-harvesting, are $Ge^{2+}$, $Sb^{3+}$ or $Bi^{3+}$ cations. While Ge-based perovskites suffer from material instability due to hydrolytic decomposition in humid environment[8–10], Sb-based perovskites show limited photovoltaic performances in spite of their good chemical stability[8]. Bismuth is a heavy metal with a stable oxidation state, little toxicity, and a similar polarizable electron density than Pb[11]. Thanks to the formation of $BiX_6^{3-}$ octahedron, it can also form perovskites with diverse dimensionalities and phases, from which methylammonium bismuth iodide ($(CH_3NH_3)_3Bi_2I_9$, MBI) is at the forefront[9]. MBI is a wide-bandgap semiconductor with a zero-dimensional structure having reduced toxicity, high stability and promising photovoltaic efficiency[12]. Despite such remarkable potential, its use is limited by the low quality of the films resulted from conventional spin-coating deposition, which produces non-compact MBI layers with poor coverage and crystallinity[8]. Indeed, in the last few years, several reports have demonstrated the strong effect of solvents, precursor's concentration and processing conditions in achieving enhanced thin film coverage[13–17].

Alternative strategies to enhance film morphology combined the use of two-step processes as well as additive inclusion such as $Br^-$ or *N*-methyl-2-pyrrolidone (NMP)[18,19]. However, such procedures rely on harmful polar solvents such as DMF or DMSO, which are not feasible for mass production. In an attempt to reduce the solvent

Group for Molecular Engineering of Functional Materials, Institute of Chemical Sciences and Engineering, EPFL Valais Wallis, Rue de l'Industrie 17, 1951 Sion, Switzerland. ✉email: cristina.roldancarmona@epfl.ch; mdkhaja.nazeeruddin@epfl.ch

toxicity while improving film coverage, ethanol has been proposed as a plausible solvent for $BiI_3$, affording PCE of 0.053%[20]. Nevertheless, the toxicity is only partially reduced with this method as toxic methylamine gas is used as the organic precursor. Interestingly, additional deposition techniques like electric-field assisted spray coating[21,22], vapour-assisted solution process (VASP)[23], or chemical vapour deposition (CVD)[24,25] among others, have been also investigated, providing a record in device efficiency exceeding 3%[23]. However, having homogeneous and conformal coating are still challenging, making it difficult to employ MBI as the real alternative.

Among the different options existing in literature, thermal co-evaporation of thin film deposition is well known to produce smooth, compact and pin-hole free lead-based perovskites[26,27]. This technique offers improved process control and high reproducibility, allowing as well the processing of large area substrates. In spite of its versatility, to date, there is no report on thermal co-evaporation of MBI films. Only co-evaporated fully inorganic bismuth-based perovskites can be found in the literature, such as $A_3Bi_2I_9$ (A = Cs, Rb), or $AgBi_2I_7$[28,29].

In this work, we investigate for the first time the crystallization process of MBI thin films prepared via thermal co-evaporation of MAI and $BiI_3$. Different perovskite layers (**1**–**5**) with various MAI:$BiI_3$ ratios (conditions **1**–**5**) have been analyzed and its optoelectronic and crystallographic properties were evaluated. In order to vary MAI:$BiI_3$ ratio, the deposition rate of $BiI_3$ was kept at 0.4 (Å $s^{-1}$) while that of MAI varied from 0.4, 0.55, 0.7, 1 to 1.4 Å $s^{-1}$, obtaining the films **1**–**5** respectively. Our results confirm that by using the optimal conditions, applied to layer **4** in this manuscript, a pure phase of hexagonal MBI crystal structure with the long-term structural stability of several months is achieved. In addition, we observed the deposition of conformal MBI films with remarkable homogeneity and surface coverage, contrary to those previously reported from solution-based methodologies, paving the way towards a successful implementation into lead-free based solar cells.

## Results and discussion

MBI layers (**1**–**5**) were fabricated by thermal co-evaporation of the precursors, $BiI_3$ and methylammonium iodide (MAI), in vacuum using varying precursor ratio (conditions **1**–**5**, Table S1). For clarity, the deposited films obtained in condition **1**–**5** will be named as thin film **1**–**5**, respectively. During the co-evaporation process the deposition rate of each material was controlled by independent quartz microbalance crystals, and the materials were simultaneously heated until their sublimation temperature. We varied the film composition by fixing the rate of $BiI_3$ at 0.4 Å $s^{-1}$ while that of MAI varied from 0.4, 0.55, 0.7, 1 to 1.4 Å $s^{-1}$ (conditions **1**–**5**, respectively), and analyzed its impact on the optical properties, layer morphology and crystal structure. For comparison, thin films of pure $BiI_3$ were also fabricated and characterized.

Figure 1a shows the optical properties of the as-deposited films investigated by UV–Vis absorption spectroscopy. Whereas the $BiI_3$ film presents a characteristic sharp absorption onset at 690 nm, the gradual insertion of MAI (**1**) reduces the onset sharpness as well as the intensity in the 450–630 nm range. As observed for the layers (**2**–**5**), the absorbance spectra show an onset (~ 600 nm) and an excitonic peak (~ 500 nm) typical of MBI. Such an excitonic peak is attributed to the electron ground-excitation transition of 1S0 to 3P1 from the $Bi^{3+}$ to the face-shared bi-octahedrons $[Bi_2I_9]^{3-}$[30], thus confirming that MBI material is already formed at condition **2**. This is also validated by the absence of absorption in the 600 to 690 nm region associated to $BiI_3$, denoting a complete conversion of the precursor.

The optical bandgap of the $BiI_3$ and thin-films **1**–**5** were derived from the corresponding indirect bandgap Tauc plots (Fig. 1b) respectively, and the calculated values are presented in Table 1. An indirect bandgap of 1.80 eV ($BiI_3$), 1.81 (**1**), 2.11 (**2** and **5**), 2.12 (**3** and **4**) was obtained for the films, matching well with the reported values in literature[24,31].

To further analyze the in situ synthetized new materials, X-Ray Diffraction (XRD) measurements were employed in Bragg–Brentano geometry. Figure 1c summarizes the diffraction patterns of FTO, $BiI_3$ and the different MBI thin films (**1**–**5**). The film containing pure $BiI_3$ presents two main diffraction peaks at 12.8° and 41.6°, corresponding to the (003) and (300) lattices planes of the hexagonal crystal structure respectively (JCPDS card no. 7-269)[32]. Despite the peak at 12.8° remains in layer **1**, the incorporation of MAI gradually increases the diffraction signal at 29.2°, 31.7° and 45.5°, corresponding to (204), (205) and (4 − 2 6) lattice planes in MBI perovskite, suggesting its incipient formation. On increasing MAI content (films **2**–**3**), intense diffraction peaks centered at 14.5°, 17.1°, 29.1°, 31.7° and 32.2° are clearly visible, related to (102) (103) (204) (205) and (2 − 1 6) MBI lattice planes, thus confirming its crystallization as hexagonal phase with the P63/mmc space group proposed by Jakubas et al.[33]. The further increase in MAI ratio (film **5**) shows a decrease in the peak intensity which is concomitant to a lower absorption in the visible range (see Fig. 1a), indicative of either reduced MBI formation or lower quality. The additional peaks observed at low angle in films **4** and **5** are attributed to the MAI-excess in the film. Overall these results demonstrate promising optical and crystallographic properties for films **3** and **4**, which invites deeper characterization.

The effect of thermal co-evaporation on film morphology was evaluated by scanning electron microscopy (SEM). MBI films were deposited on top of FTO substrates under different experimental conditions. The results, illustrated in Fig. 2, demonstrate homogeneous, compact pinhole-free perovskites (Fig. 2a,c) with conformal coating along the FTO substrate (see cross-section images in Fig. 2b,d). The presence of a non-uniform, inhomogeneous surface coverage would lead to the presence of pinholes, which are the cause of short-circuits and defect sites limiting the photovoltaic performance[34]. Therefore, the presence of a complete coverage would reduce the probability of shunting by an improved separation between the electron and the hole transport layer[35]. Interestingly, in our study both conditions lead to a compact multicrystalline film, but the grain size is highly dependent on the MAI:$BiI_3$ ratios, changing from crystallites of around 50 nm for film **3**, to a crystallite size over 300 nm for film **4**. This dependence is also observed in MBI layers deposited by CVD, which grain crystal size increases twofold by increasing the MAI content in the film[36].

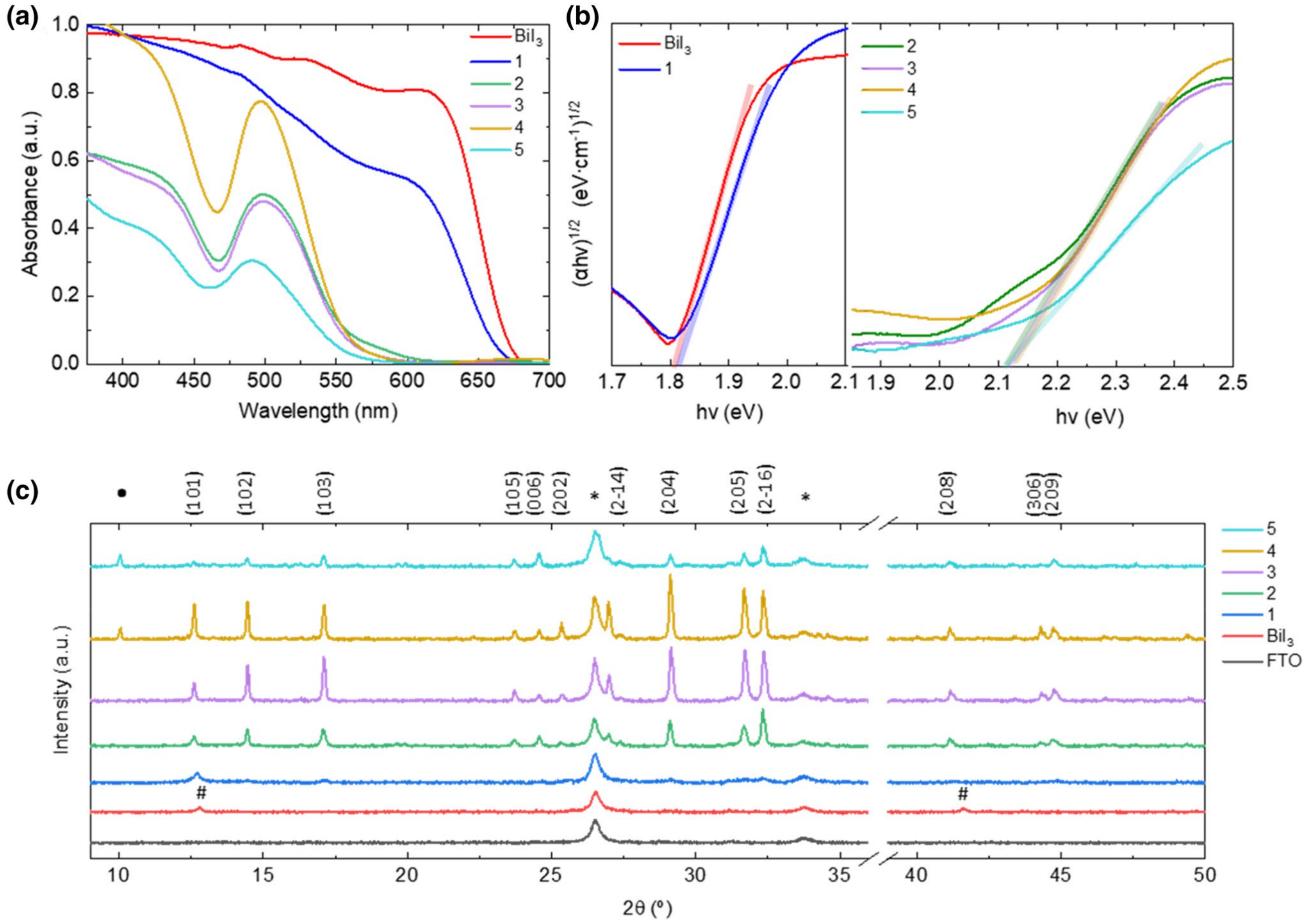

**Figure 1.** Thermally evaporated $BiI_3$ and MBI layers (**1**–**5**) with increasing MAI:$BiI_3$ ratio. (**a**) thin-film absorbance spectra, (**b**) calculated indirect Tauc plot and (**c**) thin-film X–ray diffractograms.

| Material/condition | $E_{g,indir}$ (eV) $(\alpha h\nu)^{0.5}$ [$eV^{0.5}$ $cm^{-0.5}$] |
|---|---|
| $BiI_3$ | 1.80 |
| 1 | 1.81 |
| 2 | 2.11 |
| 3 | 2.12 |
| 4 | 2.12 |
| 5 | 2.11 |

**Table 1.** Calculated indirect bandgap from their respective Tauc plots.

The long-term stability of these materials was also tested after 4 months stored under dry air at room temperature (RH ~ 10%). Figure 3 shows the evolution of the XRD patterns corresponding to the best co-evaporated films, prepared at conditions **3** and **4**, which did not present any significant change in the peak position and intensity over time. This excellent stability, observed in the co-evaporated film, is in good agreement with stoichiometric MBI films processed from the liquid phase, further demonstrating that the co-evaporation is a valid method for fabrication of pure hybrid bismuth halide perovskites. Previous studies on solution-processed MBI layers demonstrate that no or very small change is observed around the 29° diffraction peak after the first month of continuous air exposure[24,35]. Such small changes were attributed to a small amount of $Bi_2O_3$ or BiOI, formed at the MBI surface[24]. Interestingly, the peak attributed to the MAI-excess observed in film **4** disappears after aging, potentially due to its slow decomposition under 10% moisture.

The stability evolution observed for films **1**, **2** and **5** is shown in the Supporting Information. The lack of crystallographic peaks in the XRD of the as-deposited film **1** (Figure S1) denotes a high degree of an amorphous phase. However, during ageing the precursors react and the peaks attributed to MBI appears together with a peak at low angle attributed to unreacted MAI excess. Note that the film **2** (Figure S2) demonstrate the same superior stability than films **3** and **4.** However, the film with the highest MAI amount, presents the typical MBI peaks with low intensity and a peak attributed to MAI excess that disappears after ageing may be due to its slow decomposition (Figure S3).

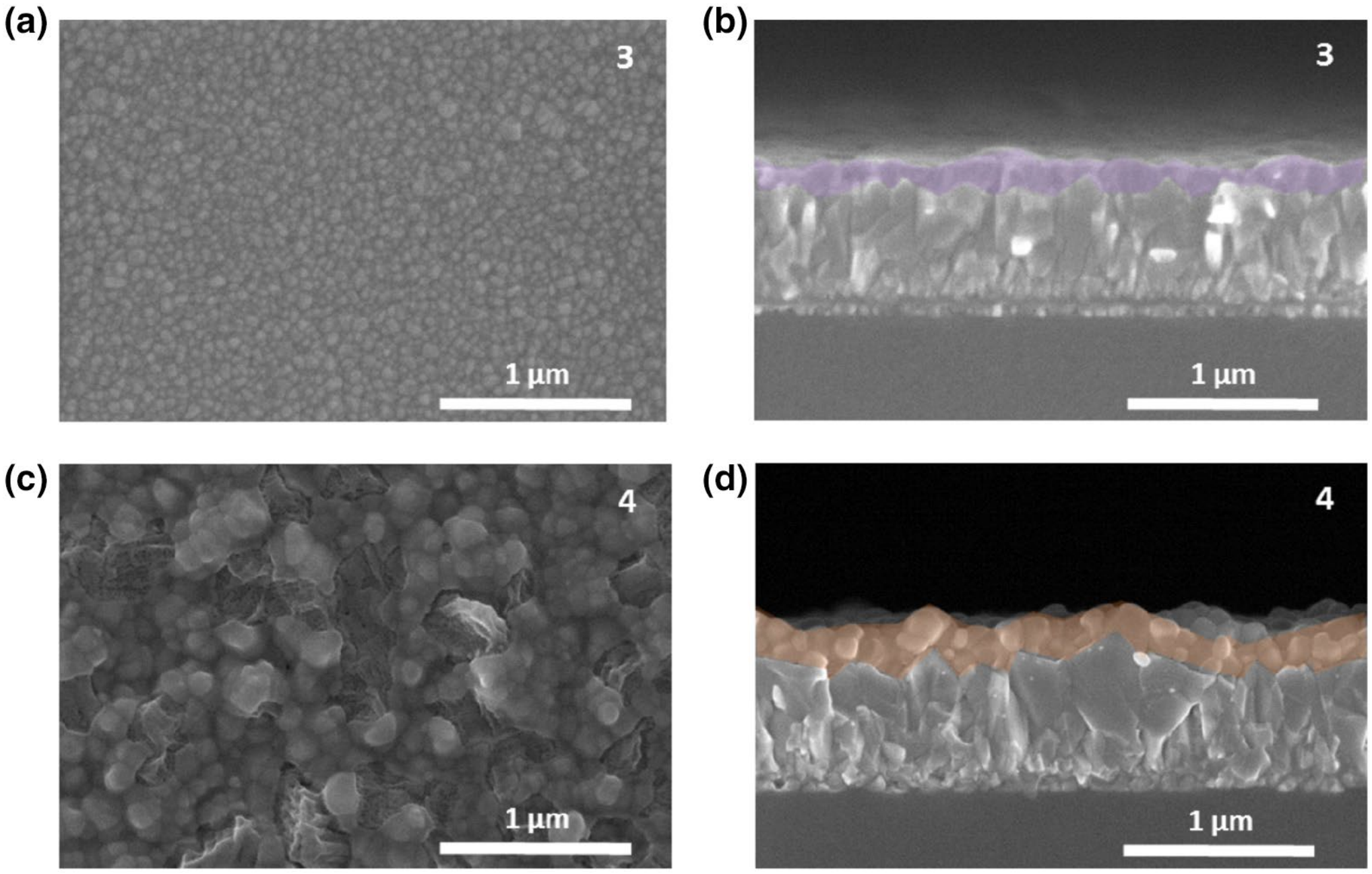

**Figure 2.** (**a**, **c**) Top-view and (**b**, **d**) cross-section scanning electron microscopy (SEM) images of co-evaporated MBI layers **3** and **4** deposited on top of FTO-coated substrates.

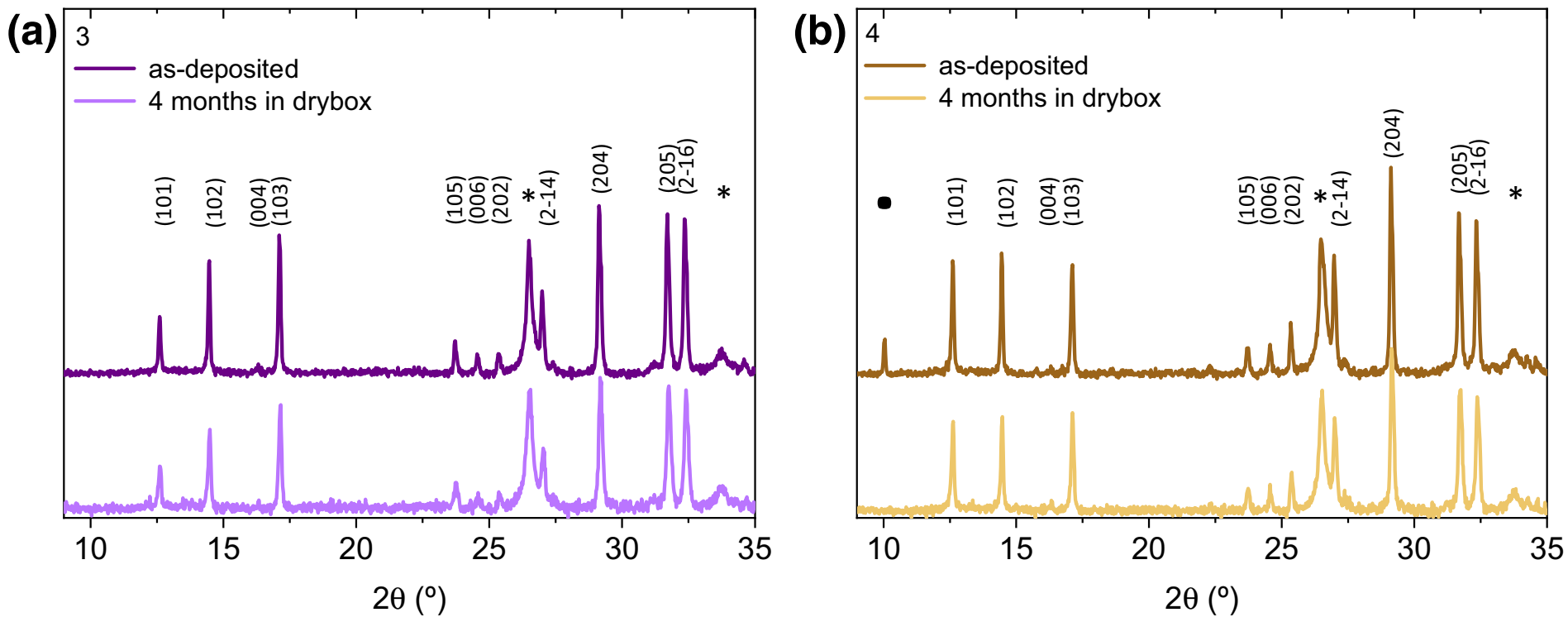

**Figure 3.** XRD patterns of: (**a**) film **3** and (**b**) film **4**, both measured just after deposition and after 4 months stored in dark, 25 °C and relative humidity (RH) of 10%. *FTO substrate and •MAI.

After the analysis of the co-evaporated films, we can conclude that the optimum thin film MBI composition is between conditions **3** and **4**. We selected film 4 for further analysis because of its film stability and adequate morphology. We further characterized the electronic properties of the MBI thin film by ultraviolet photoelectron spectroscopy (UPS), see UPS spectra in Figure S4. From the UPS measurement, the work function (WF) is calculated as the difference between the photon energy of the source, 21.22 eV, and the intercept of the secondary electron cutoff (SECO) and the zero intensity axis (Fig. 4b), and yields the value of 4.92 eV. The valence band maximum (VBM) was estimated by the intercept between the linear fitting of the valence band onset and the zero intensity axis (Fig. 4a) and is located 1.16 eV below the Fermi level, hence 6.08 eV below the vacuum level. As the calculated optical bandgap is 2.11 eV (see details below), the conduction band minimum (CBM) is therefore located at − 3.96 eV with respect to the vacuum level[37]. Finally, the Fermi energy ($E_F$) being 1.16 eV above the valence band maximum reveals the n-type character of thin film **4**. The energy level diagram is depicted in Figure S4b. These values are in the range of the values reported for solution-processed MBI films[12,38,39].

For further characterization, we verified the time-resolved photoluminescence decay with MBI film (Figure S5). The sample was deposited on the glass for this measurement. The time constant was $\tau_1 = 2.6$ ns and $\tau_2 = 13.5$ ns, which is almost corresponding to the solution-processed reports[39,40], as well as closed to bulk properties of this material[24]. From these results, we expect that the defect in the crystal could be similar to the solution-processed MBI. In addition, the PL spectra was recorded in the same sample at an excitation wavelength

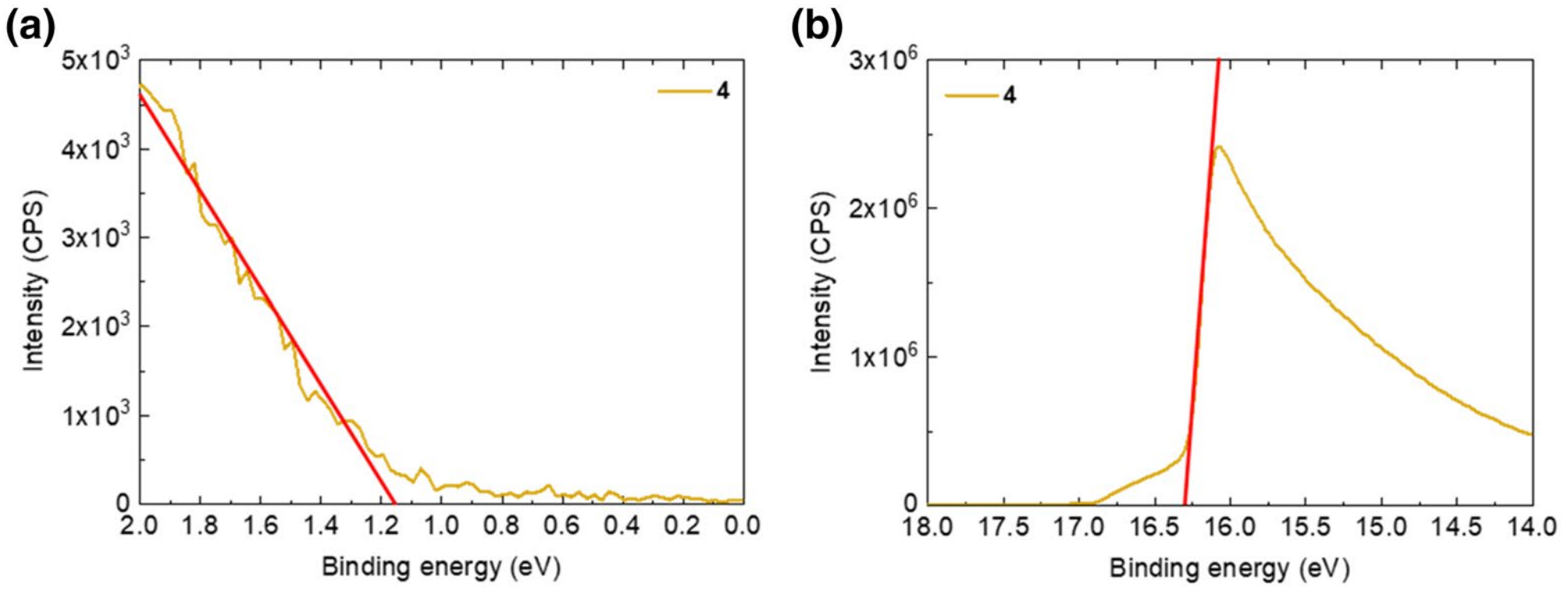

**Figure 4.** UPS measurement of MBI layer **4**. (**a**) valence band maximum and (**b**) secondary electron cutoff. The work function (WF) can be calculated as WF = *hν*-SECO, *hν* = 21.22 eV being the energy of the incident photons from a He(I) UV source.

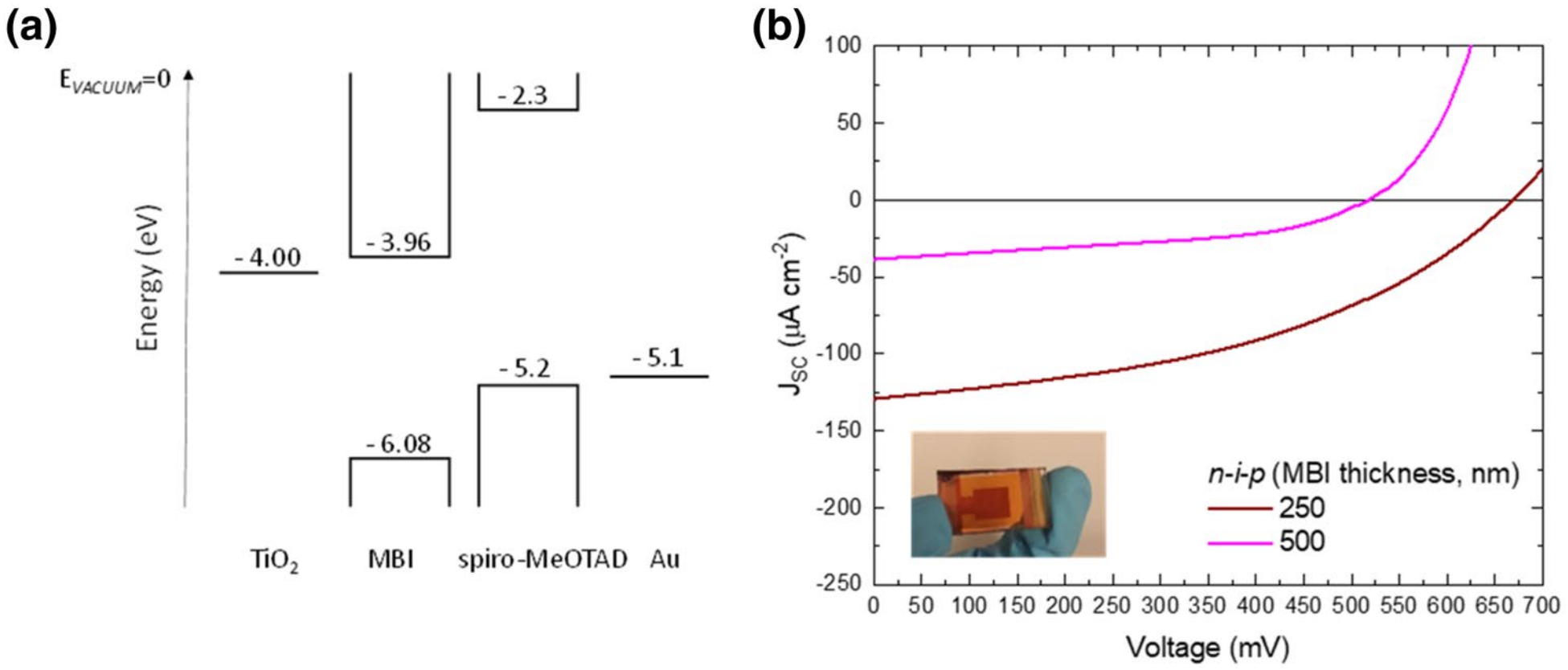

**Figure 5.** (**a**) Energy level diagram of the device layout. (**b**) *J*–*V* curves of *n-i-p* solar cells, containing 250 and 500 nm thick film co-evaporated MBI **4** as light absorber. Inset: photograph of an MBI-based perovskite solar cell device.

| Device structure | $V_{OC}$ (mV) | $J_{SC}$ (µA cm$^{-2}$) | FF | PCE (%) |
|---|---|---|---|---|
| FTO/c-$TiO_2$/MBI(250 nm)/spiro-MeOTAD/Au | 668 | 129 | 0.43 | 0.04 |
| FTO/c-$TiO_2$/MBI(500 nm)/spiro-MeOTAD/Au | 520 | 38 | 0.44 | 0.01 |

**Table 2.** Photovoltaic parameters of the best *n-i-p* solar cell containing 250 nm thick film of co-evaporated MBI layer.

of 450 nm. The PL spectra (Figure S6) presents a maximum at 581.9 nm (2.13 eV), in good agreement with the calculated band gap of the layer (2.12 eV, see Table 1). To finalize the analysis, the photovoltaic behavior of the co-evaporated MBI layer **4** was evaluated. The layer was integrated in devices with standard planar *n-i-p* architecture, sandwiched between compact $TiO_2$ acting as an electron transport layer, and doped spiro-MeOTAD as hole transport material. The energy diagram of the device layout is shown in Fig. 5a. Two different thicknesses of 250 nm and 500 nm thick MBI layers were studied and their respective *J*–*V* curves are shown in Fig. 5b. The photovoltaic parameters extracted from the curves are presented in Table 2. The device containing 250 nm thick film presented the best solar cell performance with PCE of 0.04% ($V_{OC}$ of 668 mV, $J_{SC}$ of 129 µA cm$^{-2}$ and FF of 0.43). In contrast, the device containing 500 nm MBI layer presented lower photovoltaic parameters, which we attribute to increased charge recombination due to the large distance carriers need to travel through the material. Note that such low performances cannot be attributed to a bad homogeneity of the film, neither to a bad materials quality. On the contrary, we believe that these values are deeply related to the high exciton binding energy of the material and the non-optimized device layout used for Bi-perovskite devices (see Fig. 5a). The external quantum efficiency (EQE) of the device containing 250 nm MBI film is depicted in Figure S7. The low

EQE values at higher-energies have attributed the loss in photogenerated carrier extraction efficiency near the MBI/spiro-MeOTAD interface, reflected also in the low $J_{SC}$ values obtained in the devices.

In summary, we report for the first time the synthesis of hybrid bismuth-based MBI semiconductor by thermal co-evaporation. After optimization of the precursor ratio (MAI and $BiI_3$), pure MBI layers can be fabricated without the need of thermal annealing. The optimized layers present the typical optical properties of MBI but with improved morphology thanks to a compact, dense and conformal coating with long-term stability. The successful deposition of MBI via co-evaporation is a revolutionary precedent for fabricating organic–inorganic bismuth-based perovskites, which may lead future investigations for optoelectronic applications. The photovoltaic performance is expected to be enhanced with the use of charge transporting layer with better energy band alignment with the MBI perovskite.

## Methods

**Materials.** Methylammonium iodide and bismuth (III) iodide anhydrous were purchased from Lumtec and TCI, respectively. FTO substrates (TEC-15) were purchased from Nippon Sheet Glass (NSG) group. Spiro-MeOTAD was provided by ChemBorum, titanium diisopropoxide bis(acetylacetonate), Li-bis(trifluoromethanesulphonyl) imide (LiTFSI) and 4-*tert*-butylpyridine (TBP) were purchased from Sigma Aldrich and FK209 Co(III) TFSI from GreatCellSolar. All of the chemicals were used as received without further purification.

**Thin-film fabrication.** FTO-coated glass substrates were cleaned by sonication in a 2% Hellmanex III solution, deionized water and ethanol for 10 min each, followed by a 15 min UV–$O_3$ treatment. After the surface treatment, the substrates were introduced into the vacuum chamber for perovskite deposition. The perovskite layer was deposited at the studied precursor evaporation rates **1**–**5** (Table S1). The deposition was made in the PRO Line PVD 75 vacuum chamber from Kurt J. Lesker Company equipped with thermal evaporator sources. The thin films were fabricated by thermal co-evaporation of the precursors bismuth iodide ($BiI_3$) and methylammonium iodide (MAI) in vacuum. During co-evaporation, the deposition rate of each material was controlled by independent quartz microbalance crystal sensors (QCM). MAI and $BiI_3$ were simultaneously heated under high vacuum until its sublimation temperature. To achieve a high degree of uniformity, the substrates were kept at room temperature and were continuously rotating during the deposition. To vary the film composition, the deposition rate of $BiI_3$ was kept at 0.4 (Å $s^{-1}$) while that of MAI varied from 0.4, 0.55, 0.7, 1 to 1.4 Å $s^{-1}$, obtaining the films **1**–**5** respectively (Table S1). The final thickness of MBI layer was around 250 nm.

**Thin-film characterization.** UV–Vis spectra were measured by using Lambda 950S spectrometer (PerkinElmer, Inc.). X-ray diffractograms were recorded in Bragg–Brentano geometry on a Bruker D8 Advance equipped with a ceramic tube (Cu anode, $\lambda = 1.54060$ Å). Scanning electron microscopy (SEM) images were recorded by in-lens detector of FEI Teneo Schottky Field Emission SEM at tension of 5 kV. The UPS measurements were carried out on an Axis Supra (Kratos Analytical) using UV from an He(I) source. The pass energy of the analyser was set to 10 eV. The samples were electrically grounded, and the photoelectron intensity is presented as a function of the binding energy referenced at the Fermi level of the analyser. Time-resolved photoluminescence decay was measured in a sample of MBI deposited on glass with Flurolog TCSPC (FL-1000, HORIBA) proved at 575 nm upon excitation at 455 nm. The photoluminescence emission was measured in a sample of MBI deposited on glass with a fluorescence spectrometer PerkinElmer LS 55 upon excitation at 450 nm. The thin-film stability test was performed by keeping the layers without encapsulation in dark at dry air (10% humidity) at 25 °C.

**Device fabrication.** FTO glass were chemically etched with Zn powder and an HCl solution. The chemically etched substrate was cleaned by sonication in a 2% Hellmanex solution, deionized water and ethanol for 10 min each, followed by a 15 min UV–$O_3$ treatment. Titanium diisopropoxide bis(acetylacetonate) (TAA) in ethanol solution (1:15 v/v) was sprayed at 450 °C and kept annealing during 30 min. Once the substrates were cooled down, they were introduced into the vacuum chamber for perovskite deposition. The perovskite layers were deposited until the desired thickness of 250 or 500 nm, respectively. After that, the spiro-MeOTAD solution (60 mM) was spin-coated at 4,000 rpm for 30 s. The molar ratio of additives for spiro-MeOTAD solution was 0.5, 3.3 and 0.05 for Li-TFSI, TBP and FK209, respectively. Finally, a 80 nm thick Au was thermally evaporated in vacuum ($10^{-6}$ mbar).

**Device characterization.** *J*–*V* curves were measured by using a 2,400 Keithley system in a scan rate of 50 mV $s^{-1}$ and 10 mV voltage step in combination with a Xe-lamp Oriel sol3A sun simulator (Newport Corporation), which was calibrated to AM1.5G standard conditions by using an Oriel 91,150 V reference cell. The solar cells were measured without light soaking and with an illumination area through a shadow mask of 16 $mm^2$. EQE was measured with IQE200B Quantum Efficiency Measurement System (Oriel, Newport).

## Acknowledgements

This Project has received funding from the European Union's Horizon 2020 research and innovaton programme under the Marie Skłodowska–Curie Grant Agreement No. 754462. The authors thank to the project German Research Foundation (DFG) (Projekt number 424101351)-Swiss National Foundation (SNF) (200021E_186390). C. M. thanks to the co-funded fellowship Horizon 2020-Marie Skłodowska–Curie (Agreement No. 754462).

## Author contributions

C.M. conceived the ideas, deposited and characterized films and devices and wrote the manuscript. A.A.S. and H.K. measured steady-state and time-resolved photoluminescence. M.M. measured ultraviolet photoelectron spectroscopy. M.K.N. and C.R.-C. supervised the project.

## Competing interests

The authors declare no competing interests.

## Additional information

**Supplementary information** is available for this paper at https://doi.org/10.1038/s41598-020-67606-1.

**Correspondence** and requests for materials should be addressed to C.R.-C. or M.K.N.